\documentclass[aps,prd,reprint,groupedaddress]{revtex4-1}

\usepackage{graphicx}
\usepackage{amssymb, amsmath,mathtools,amsthm}
\newtheoremstyle{dotless}{}{}{\itshape}{}{\bfseries}{}{ }{}
\theoremstyle{dotless}
\makeatletter
\def\@endtheorem{\endtrivlist}
\makeatother
\newtheorem*{proposition*}{}

\newcommand{\be}{\begin{equation}}
\newcommand{\bea}{\begin{align}}
\newcommand{\eea}{\end{align}}
\newcommand{\beq}{\begin{equation}}
\newcommand{\ee}{\end{equation}}
\newcommand{\eeq}{\end{equation}}

\newcommand{\mnras}{{MNRAS}}

\newcommand{\ssr}{{Space~Sci.~Rev.}}

\newcommand{\physrep}{{Phys.~Rep.}}
\newcommand{\physscr}{{Phys.~Scr}}

\begin{document}


\title{Black hole Meissner effect and entanglement}


\author{Robert F. Penna}
\email[]{rpenna@mit.edu}
\affiliation{Department of Physics and Kavli Institute for Astrophysics and Space Research,
Massachusetts Institute of Technology, Cambridge, Massachusetts 02139, USA}


\date{\today}

\begin{abstract}

Extremal black holes tend to expel magnetic and electric fields.  Fields are unable to reach the horizon because the length of the black hole throat blows up in the extremal limit.  The length of the throat is related to the amount of entanglement between modes on either side of the horizon.  So it is natural to try to relate the black hole Meissner effect to entanglement.  We derive the black hole Meissner effect directly from the low temperature limit of two-point functions in the Hartle-Hawking vacuum.  Then we discuss several new examples of the black hole Meissner effect, its applications to astrophysics, and its relationship to gauge invariance.

\end{abstract}


\maketitle

\section{Introduction}
\label{sec:intro}

The prototypical example of the black hole Meissner effect was found 40 years ago.  If a Kerr black hole is immersed in a uniform magnetic field aligned with its spin axis, and spun up to its extremal limit, then the flux of the field through any patch of the horizon drops to zero \cite{1975PhRvD..12.3037K,1985MNRAS.212..899B}.  Reissner-N\"{o}rdstrom \cite{1980PhRvD..22.2933B}, Kerr-Newman  \cite{bicak1989,1991JMP....32..714K,2000PhyS...61..253K,2014PhRvD..89d4029G}, and string theory black holes \cite{1998PhRvD..58h4009C} all show the same behavior. Electric fields are also expelled \cite{2004NCimB.119..785R,2008PhRvD..77f4020B}.  

The effect is reminiscent of the Meissner effect of superconductors.  Superconductors expel magnetic fields as they are cooled.  Extremal black holes have vanishing Hawking temperature, so black holes also tend to expel magnetic fields as they are cooled.

This has generated interest for its astrophysical applications.  The standard model of black hole jets \cite{bz77} requires magnetic fields threading the horizon \cite{2014arXiv1403.0938P}.  The fields torque the horizon and extract its rotational energy  (according to the membrane paradigm \cite{1986bhmp.book.....T,2013MNRAS.436.3741P}).  So if the Meissner effect expels the field, then the jets will be quenched \cite{1991NYASA.631..235P,2000NCimB.115..739B,2007IAUS..238..139B,2014arXiv1403.0938P}.  There are field geometries which evade the Meissner effect and may power jets from extremal black holes \cite{1985MNRAS.212..899B,2007MNRAS.377L..49K,2014arXiv1403.0938P}. 

We would like to understand the origin of the black hole Meissner effect.  The geometrical explanation is based on the fact that extremal black holes have infinitely long throats \cite{1986bhmp.book.....T,2014arXiv1403.0938P}.  The proper spacelike distance between the horizon and points outside the horizon becomes infinite in the extremal limit.  Electric and magnetic fields fall off with distance.  So if stationary charges and currents are placed outside the horizon, then the fields will vanish at the horizon in the extremal limit.  (The distance to the horizon remains finite in timelike directions, so infalling observers can cross the horizon in finite proper time \cite{1972ApJ...178..347B}.)  This is the Meissner effect.

The length of the throat is related to the entanglement between modes on either side of the horizon.  For example, consider a scalar field with mass $m$.  The two-point correlation function for spacelike separations is 
\beq\label{eq:corr}
\langle \phi(x)\phi(x')\rangle \sim e^{-mL},
\eeq
where $L$ is the length of the shortest geodesic connecting $x$ and $x'$, and we have assumed $m\gg 1/L$.    One usually views the metric as fundamental and uses eq. \eqref{eq:corr} to compute correlations.  Correlations fall off with distance. However, sometimes the opposite perspective is useful.  For example, black holes in Anti de Sitter (AdS) space are related to entangled states of conformal field theories (CFTs) in one less dimension \cite{2003JHEP...04..021M}.  One can use eq. \eqref{eq:corr} to compute the length of the emergent dimension  from the entanglement of the CFTs (e.g., \cite{2000PhRvD..61d4007B,2000PhRvD..62d4041L,2003PhRvD..67l4022K,2014JHEP...03..097H,2010GReGr..42.2323V,Andrade:2013rra,Leichenauer:2014nxa}).
Unentangling the CFTs causes the length of the emergent dimension  to grow.  Even in spacetimes which are not described by the AdS/CFT correspondence, there are close relationships between geometry and entanglement.  So it is natural to try to understand the relationship between the black hole Meissner effect and entanglement.

Our main result is to derive the black hole Meissner effect directly from the properties of the Hartle-Hawking vacuum \cite{1976PhRvD..13.2188H,1976PhLA...57..107I}.  Correlators in this vacuum satisfy a Kubo-Martin-Schwinger (KMS) condition \cite{doi:10.1143/JPSJ.12.570,PhysRev.115.1342,1996IJMPB..10.1755T} which implies that modes on either side of the horizon become unentangled at low Hawking-Unruh temperatures.  This makes it impossible for charges and currents in the exterior to create flux on the horizon. This gives the black hole Meissner effect.  
This approach to the problem clarifies why the Meissner effect appears at low Hawking-Unruh temperatures.  It also clarifies why the effect is a universal feature of extremal black holes, and it connects the effect to similar features of Rindler and de Sitter horizons.

Given the similarities between the black hole Meissner effect and the ordinary Meissner effect of superconductors, it is tempting to try to find a common underlying cause.  The ordinary Meissner effect is caused by gauge symmetry breaking.  The electromagnetic field inside the superconductor acquires a mass and it becomes energetically favorable for the field to be expelled.  Black hole horizons break gauge invariance in a certain sense \cite{1995IJMPA..10.1969B,1996NuPhB.461..581B,2008PhLB..670..141B,2012PhRvD..85h5004D}.  However, as we discuss in \S\ref{sec:discuss}, the role this plays in the black hole Meissner effect appears to be the reverse of the role it plays in the ordinary Meissner effect.

The ordinary Meissner effect may be described phenomenologically using London's equation.  Chamblin et al. \cite{1998PhRvD..58h4009C} defined a current on black hole horizons and showed that it obeys a version of London's equation.  The horizon's ``London penetration depth'' can remain nonzero in the extremal limit, but horizons have zero thickness, so it is not entirely clear from this perspective why the field should be expelled. 

Our paper is organized as follows.  In \S\ref{sec:meissner}, we derive the black hole Meissner effect 
from the KMS property of the Hartle-Hawking vacuum.    In \S\ref{sec:discuss}, we discuss new examples of the black hole Meissner effect, its applications to astrophysics, and its relationship to gauge invariance.  We summarize and conclude in \S\ref{sec:conc}.

\section{Black hole Meissner effect from entanglement}
\label{sec:meissner}

\subsection{Hartle-Hawking vacuum}

\begin{figure}
\includegraphics[width=\columnwidth]{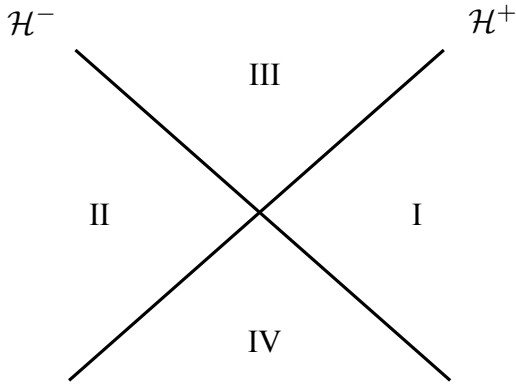}
\caption{A bifurcate Killing horizon, $\mathcal{H}=\mathcal{H}^-\cup \mathcal{H}^+$, divides spacetime into four regions. For example, for a Schwarzschild black hole, regions I and II are the right and left exterior regions, region III is the black hole interior, and region IV is the white hole interior.}
\label{fig:penrose}
\end{figure}

Consider a stationary spacetime with Killing vector $\xi$ and bifurcate Killing horizon $\mathcal{H}=\mathcal{H}^-\cup\mathcal{H}^+$. The Killing horizon divides spacetime into four regions (see fig. \ref{fig:penrose}).  For example, for the eternal Schwarzschild black hole, regions I and II are the right and left exterior regions, region III is the black hole interior, and region IV is the white hole interior.  A black hole formed from stellar collapse does not have regions II and IV but including them makes it easier to describe the vacuum.

Assume fields are in the Hartle-Hawking vacuum state \cite{1976PhRvD..13.2188H,1976PhLA...57..107I}.  This vacuum does not always exist \cite{1991PhR...207...49K}.  For example, there is no Hartle-Hawking vacuum for asymptotically flat Kerr black holes \footnote{this is related to the fact that the metric does not have a globally defined timelike Killing vector}.  However, it is possible to arrange the Hartle-Hawking vacuum in this case by surrounding the black hole with a mirror \cite{Duffy:2005mz,Frolov:1989jh}.  Black holes formed from stellar collapse will not be in the Hartle-Hawking state, but consider it as a toy model for the true vacuum.

In this state, a stationary observer in region I perceives a thermal bath of Hawking-Unruh particles with inverse temperature $\beta=2\pi |\xi|/\kappa$, where $\kappa$ is the surface gravity of the horizon.  The extremal limit is $\beta\rightarrow \infty$.  For a Kerr black hole, this is the limit of extremal spin.

Consider a free Maxwell field on this spacetime.  Let $A_\mu(t,x)$ be the vector potential in region I and let $\tilde{A}_\mu(t,x)$ be the vector potential in region II. In the Hartle-Hawking state, $|\Psi\rangle$, they are related by a shift in imaginary time \cite{1996IJMPB..10.1755T,Ojima19811}:
\begin{align}\label{eq:kms}
 \tilde{A}^\dagger_\mu(t,x) |\Psi\rangle &=  A_\mu(t+i \beta(x)/2,x)|\Psi\rangle,\\
\langle \Psi | \tilde{A}^\dagger_\mu(t,x) &= \langle \Psi | A_\mu(t-i \beta(x)/2,x).\label{eq:kms2}
\end{align}
Combining these relationships (and using the fact that $A_\mu$ and $\tilde{A}_\mu^\dagger$ commute) gives the KMS condition,
\begin{align}\label{eq:kms3}
\langle \Psi | &A_\mu(t,x) A_\nu(t',x') |\Psi\rangle \notag\\
&= \langle \Psi | A_\nu(t',x') A_\mu(t+i\beta(x),x)   |\Psi\rangle.
\end{align}
These relationships characterize the thermal nature of the
 Hartle-Hawking state.  They are valid up to gauge transformations.  We have followed the conventions of \cite{Ojima19811}.

\subsection{Unentangling regions I and II}

Consider 2-point correlations between regions I and II in the Hartle-Hawking vacuum (see fig. \ref{fig:penrose}). Our goal is to show that they vanish in the extremal limit $\beta\rightarrow \infty$.   Spatial variables are suppressed throughout this section. 

The causal Green's function is defined by  \cite{1996IJMPB..10.1755T,Ojima19811}
\begin{align}
\langle \Psi |& T\tilde{A}^\dagger_\mu (t)A_\nu(t') |\Psi \rangle\notag\\
&=\theta(t-t') \langle \Psi | \tilde{A}^\dagger_\mu (t)A_\nu(t') |\Psi\rangle \notag\\
&+\theta(t'-t) \langle \Psi | A_\nu(t') \tilde{A}^\dagger_\mu (t) |\Psi \rangle\notag\\
&=\frac{i}{2\pi}\int_{-\infty}^\infty dp_0e^{-ip_0(t-t')}G_{\mu\nu}(p_0).\label{eq:agreen}
\end{align}
Applying the shift conditions \eqref{eq:kms} and \eqref{eq:kms2} gives
\begin{align}
\langle \Psi | &T\tilde{A}^\dagger_\mu (t)A_\nu(t') |\Psi \rangle\notag\\
&=\theta(t-t') \langle \Psi | A_\mu (t-i\beta/2)A_\nu(t') |\Psi \rangle\notag\\
&+\theta(t'-t) \langle \Psi | A_\nu(t') A_\mu (t+i\beta/2) |\Psi \rangle.
\end{align}
The KMS relation \eqref{eq:kms3} gives
\begin{align}
\langle \Psi | &T\tilde{A}^\dagger_\mu (t)A_\nu(t') |\Psi \rangle\notag\\
&=\theta(t-t') \langle \Psi | A_\mu (t-i\beta/2)A_\nu(t') |\Psi \rangle\notag\\
&+\theta(t'-t) \langle \Psi | A_\mu (t-i\beta/2) A_\nu(t')  |\Psi \rangle.
\end{align}
The two terms on the rhs are the same up to factors of $\theta$.  So we have simply
\beq
\langle \Psi | T\tilde{A}^\dagger_\mu (t)A_\nu(t') |\Psi \rangle
=\langle\Psi | A_\mu (t-i\beta/2) A_\nu(t')  |\Psi \rangle.
\eeq

Now introduce the Fourier representation 
\begin{align}
\langle \Psi | A_\mu (t)A_\nu(t') |\Psi \rangle =\frac{1}{2\pi}\int_{-\infty}^\infty dp_0 e^{-ip_0(t-t')}I_{\mu\nu}(p_0).
\end{align}
Plugging into the previous equation gives
\begin{align}
\langle \Psi | &T\tilde{A}^\dagger_\mu (t)A_\nu(t') |\Psi \rangle\notag\\
&=\frac{1}{2\pi}\int_{-\infty}^\infty dp_0 e^{-ip_0(t-t')}e^{-\beta p_0/2}I_{\mu\nu}(p_0).
\end{align}
Comparing with \eqref{eq:agreen} gives
\beq\label{eq:agreen0}
G_{\mu\nu}(p_0) = i e^{-\beta p_0/2}I_{\mu\nu}(p_0).
\eeq
For physically realistic $I_{\mu\nu}$, we may restrict attention to finite $p_0>0$.  The extremal limit $\beta\rightarrow \infty$ of \eqref{eq:agreen0}, together with \eqref{eq:agreen}, then gives
\beq
\langle \Psi | T\tilde{A}^\dagger_\mu (t)A_\nu(t') |\Psi \rangle = G_{\mu\nu}(p_0) =  0.
\eeq
In other words, correlations between regions I and II vanish in the extremal limit $\beta\rightarrow \infty$, as claimed.  We will now argue that this causes the black hole Meissner effect.

\subsection{Black hole Meissner effect}

Restoring spatial variables, we have that as $\beta\rightarrow \infty$,
\beq\label{eq:aa0}
\langle \Psi | T \tilde{A}^\dagger_\mu (t,x)A_\nu(t',x') |\Psi \rangle \rightarrow 0.
\eeq
Now take the limit in which the point in region II approaches the horizon.  A stationary observer at the horizon perceives $\beta=0$, so $\tilde{A}_\mu^\dagger$ and $A_\mu$ match at the horizon by \eqref{eq:kms}-\eqref{eq:kms2}. We therefore conclude from \eqref{eq:aa0} that correlations in region I between points on the horizon and points not on the horizon vanish in the extremal limit.  In other words,
\beq\label{eq:2point0}
\langle \Psi | T A_\mu (t,x_+)A_\nu(t',x') |\Psi \rangle \rightarrow 0,
\eeq 
where $x_+$ is on the horizon and $x'$ is not on the horizon.   We will now argue that this causes the black hole Meissner effect.

Consider a stationary, localized distribution of charges and currents, $j^\mu$, in region I.  
The vector potential sourced by the current is
\beq\label{eq:green}
A_\mu(x) = \int d^3x' \Delta_{\mu\nu}(x;x') j^\nu(x'),
\eeq
where $\Delta_{\mu\nu}(x;x')$ is the Green's function for Maxwell's equations.  The time-dependence has dropped out because everything is stationary.    The Green's function for Maxwell's equations is the same as the two-point function of a free Maxwell field.  Assume the current is sufficiently weak that it does not backreact significantly on the metric, so the two-point function may be evaluated in the Hartle-Hawking vacuum.  

Now take the extremal limit.  Eqs. \eqref{eq:2point0} and \eqref{eq:green} imply that the vector potential vanishes on the horizon:
\beq\label{eq:meissner}
A_\mu(x_+)\rightarrow0,
\eeq
up to gauge transformations.  The magnetic flux threading a patch, $\Sigma$, of the horizon is 
\beq\label{eq:flux}
\int_{\Sigma}dA = \int_{\partial\Sigma} A = 0,
\eeq
by Stokes' theorem.  So the flux of the field through any patch of the horizon vanishes in the extremal limit.  This is the Meissner effect for magnetic fields.  
Modes on either side of the horizon become unentangled as $\beta\rightarrow \infty$ and it is impossible for currents in the exterior to create flux on the horizon.

The geometrical interpretation of the black hole Meissner effect relies on the fact that the length of the throat blows up in the extremal limit.  This is the geometrical manifestation of \eqref{eq:2point0}.  We have derived this fact from the KMS property of the Hartle-Hawking vacuum at low Hawking-Unruh temperatures.

It is straightforward to extend this argument to electric fields.  In place of \eqref{eq:2point0}, consider the 2-point function
\beq
\langle \Psi | TE_i(t,x) A_\mu(t',x') | \Psi \rangle,
\eeq
where $E_i$ is the electric field. The same reasoning implies that this 2-point function vanishes in the extremal limit when $x$ is on the horizon (and $x'$ is not also on the horizon).  So it is impossible for charges in the exterior to create electric fields at the horizon.  This gives the black hole Meissner effect for electric fields.

\section{Discussion}
\label{sec:discuss}

\subsection{Examples}

The discussion in section \ref{sec:meissner}  applies to all extremal black holes.  In fact, it suggests a natural extension of the black hole Meissner effect to non-black hole spacetimes.

The simplest example of a stationary spacetime with bifurcate Killing horizon is Rindler space, the causal patch of a uniformly accelerating observer in Minkowski space.  A Rindler observer with acceleration $a$ perceives a thermal bath with inverse temperature $\beta = 2\pi/a$.  In the usual Rindler coordinates, the separation between observer and horizon is $z=1/a$.  The horizon is an infinite plane.

Now consider a stationary (with respect to Rindler time) distribution of charges and currents, $j^\mu$.  Consider the electric and magnetic fluxes threading the horizon.  Cooling the Rindler vacuum is the same as decelerating $j^\mu$.  In the limit $\beta\rightarrow \infty$, the distance to the horizon, $z=1/a$, is infinite and the fluxes through any finite patch of the horizon go to zero.  So the Rindler space Meissner effect is  simply a restatement of the fact that electric and magnetic fields generated by localized sources go to zero at infinity. 

A subtlety of this example is that the total electric and magnetic fluxes threading the Rindler horizon generally remain nonzero because the horizon is an infinite plane.  The fluxes only vanish on finite patches of the horizon.

The next simplest example comes from turning on a cosmological constant.  In de Sitter space, even non-accelerating observers perceive a thermal vacuum.  The inverse temperature is $\beta = \ell/2\pi$, where $\ell=\sqrt{3/\Lambda}$ is the radius of the de Sitter horizon.  In the $\beta\rightarrow \infty$ limit, the radius of the horizon blows up.  Again, as in Rindler space, the electric and magnetic fluxes through any finite patch of the horizon drop to zero in this limit.

Finally, for completeness, suppose we instead turn on a negative cosmological constant.  In Anti de Sitter space, there is a minimum acceleration, $a=\sqrt{-\Lambda/3}$, before an observer perceives a thermal vacuum \cite{Deser:1997ri,Parikh:2012kg}.  At this critical acceleration, the inverse temperature is $\beta=\infty$ and the distance to the Rindler horizon is infinite.  So a version of the black hole Meissner effect also exists for Rindler observers in Anti de Sitter space.

These examples are a useful consistency check of our formulation of the Meissner effect and they illustrate the interplay between entanglement and geometry: sending $\beta\rightarrow \infty$ always causes the distance to the horizon to blow up.   
Of course, the situation for extremal black holes is more interesting.  In black hole spacetimes it is much less obvious why and how the distance to the horizon should blow up  and why this should cause the Meissner effect. Our derivation of the black hole Meissner effect ties the Rindler, (Anti) de Sitter, and black hole examples together.  They all have a common origin in the entanglement of the vacuum.

\subsection{Evading the Meissner effect}
\label{sec:evade}

If there was no way to evade the black hole Meissner effect, then jets from extremal black holes would be quenched.  
The standard model of spin-powered jets \cite{bz77} requires magnetic fields threading the horizon \cite{2014arXiv1403.0938P}.  The jets are powered by magnetic torques acting on the horizon \cite{1986bhmp.book.....T,2013MNRAS.436.3741P}.
There is observational evidence for near-extremal Kerr black holes in our galaxy  \cite{2006ApJ...652..518M,2011ApJ...742...85G,2013SSRv..tmp...73M}.  So it is important to understand how the Meissner effect can be evaded.  

Our derivation assumes a stationary distribution of charges and currents.  There are examples of non-stationary fields which thread extremal horizons.  For example, a tilted field may thread the extremal Kerr horizon \cite{1985MNRAS.212..899B}.

We considered charges and currents outside the horizon and then took the $\beta\rightarrow \infty$ limit.  This precludes the possibility of charges and currents sitting exactly on the horizon, because the inverse Hawking-Unruh temperature there is always $\beta=0$. Charges and currents sitting on the horizon can continue to thread the horizon with flux in the extremal limit.  

Actually, in the extremal limit, the black hole develops an infinite throat \cite{1972ApJ...178..347B}.  In Boyer-Lindquist coordinates, the entire throat is at the radius of the horizon, $r=r_+$.  So charges and currents  in the throat may thread the extremal horizon with flux.   

To power jets in the usual way, the field must not only thread the horizon, but must also extend to infinity where it can dissipate its energy in a load region.  Fields which become radial in the throat (such as split-monopoles) are able to do this \cite{2014arXiv1403.0938P}.

\subsection{Gauge invariance}
\label{sec:gauge}

Given the similarities between the black hole Meissner effect and the ordinary Meissner effect of superconductors, it is tempting to try to find a common origin.  The ordinary Meissner effect is caused by gauge symmetry breaking.  So one might expect the black hole Meissner effect is also related to gauge symmetry breaking.  In this section we argue that it is, but not in the same way as ordinary superconductors.

Horizons break gauge symmetries in a certain sense \cite{1995IJMPA..10.1969B,1996NuPhB.461..581B,2008PhLB..670..141B,2012PhRvD..85h5004D}.  
Suppose one treats region I of fig. \ref{fig:penrose} as a spacetime in its own right and quantizes the electromagnetic field in this region.  This is natural from the perspective of observers who remain outside the horizon because no signals can reach them  from the interior (at least semiclassically).   However, it creates several problems.  Charges can flow across the horizon and disappear, so charge conservation is broken.  Field lines can  terminate at the horizon, so the Gauss constraint is broken.   

To fix these problems, one can endow the horizon with fictitious surface charges and currents.  Classically, these are the surface charges and currents of the black hole membrane paradigm \cite{1986bhmp.book.....T,1998PhRvD..58f4011P}.  The membrane charge densities are defined as the normal components of the electric and magnetic fields at the horizon:
\begin{align}
\sigma_H &= E^n,\\
\tilde{\sigma}_H &= B^n.
\end{align}
The black hole Meissner effect says $\sigma_H=\tilde{\sigma}_H=0$ for stationary fields in the extremal limit.

This discussion may be rephrased by saying that the horizon breaks gauge invariance.  To see this more clearly, consider a closed Wilson loop threading the horizon.  Restricting the loop to region I gives an open Wilson line that begins and ends on the horizon, at points $p$ and $q$, say.  Under a gauge transformation $A \rightarrow A+d\epsilon$, the Wilson line transforms as
\begin{align}
&\exp\left[-i\int_p^q dx^\mu A_\mu\right] \notag\\
&\rightarrow e^{-ie [\epsilon(q)-\epsilon(p) ]} \exp\left[-i\int_p^q dx^\mu A_\mu\right].
\end{align}
This is invariant only if $\epsilon(q)=\epsilon(p)$.  So the theory restricted to region I is only invariant under gauge transformations which are constant on the horizon.  It is in this sense that the horizon breaks gauge invariance.

One can restore gauge invariance by introducing ``edge states'' on the horizon  \cite{1995IJMPA..10.1969B,1996NuPhB.461..581B,2008PhLB..670..141B,2012PhRvD..85h5004D}.  (Including the edge states is necessary if, for example, one wants to compute the entanglement entropy between modes on either side of the horizon.)
The edge states are defined so that the full theory of edge states and exterior states is gauge invariant.   In the classical limit, the edge states are the charges and currents of the black hole membrane paradigm.

Now consider the extremal limit.  Equation \eqref{eq:meissner} says that stationary gauge fields vanish at extremal horizons (up to gauge transformations).  So in the extremal limit, Wilson loops threading the horizon may be restricted to Wilson loops lying entirely in region I without ambiguity.  In this sense, extremal horizons do not break gauge invariance for stationary fields.  There is no need for edge states, and $\sigma_H=\tilde{\sigma}_H=0$, just as there is no need for edge states at asymptotic infinity.

So the black hole Meissner effect is related to gauge symmetry breaking, but the effect is the reverse of what happens in superconductors.   Fields are expelled from superconductors because gauge invariance is broken, but they are expelled from extremal horizons because (in a certain sense) gauge invariance is restored.

\section{Conclusions}
\label{sec:conc}

A number of examples have been found in which extremal black holes expel magnetic and electric fields.  We have given an overarching explanation for the phenomenon, in which the black hole Meissner effect follows directly from black hole thermodynamics.  The effect is a direct consequence of the low Hawking-Unruh temperature behavior of thermal correlators.  Modes on either side of the horizon become unentangled at low Hawking-Unruh temperatures, and it becomes impossible for charges and currents in the exterior to source fields on the horizon.

Framed in this language, it is natural to extend the effect to non-black hole spacetimes.  In particular, Rindler and de Sitter horizons also exhibit a version of the effect.  Lowering the Hawking-Unruh temperature in Rindler and de Sitter spacetimes causes the horizons to move to infinity.  Electric and magnetic fluxes threading finite patches of the horizons drop to zero.  In these examples, the Meissner effect is simply a restatement of the fact that fields vanish at infinity.  These examples provide useful consistency checks of our method.  They also give simplified models for understanding the original Meissner effect in extremal black hole spacetimes.

It is tempting to speculate that the black hole Meissner effect is, in some sense, the ordinary Meissner effect of superconductors in disguise.  Indeed, as we discussed in \S\ref{sec:gauge}, horizons break gauge invariance in a certain sense.  However, we have argued that the role played by gauge invariance is in some sense the reverse of the role it plays in the ordinary Meissner effect.

Our interest in the Meissner effect was originally motivated by the astrophysical problem of understanding black hole jets.  The Meissner effect can quench jets, so it is important to understand what causes it and how it can be evaded.  Our explanation of the Meissner effect gives a new way of understanding this problem.

\begin{acknowledgments}
We thank H. Liu and J. Sonner for discussions and R. Emparan for correspondence.  R.F.P is supported by a Pappalardo Fellowship in Physics at MIT. 
\end{acknowledgments}

\bibliography{msnew}

\begin{thebibliography}{47}%
\makeatletter
\providecommand \@ifxundefined [1]{%
 \@ifx{#1\undefined}
}%
\providecommand \@ifnum [1]{%
 \ifnum #1\expandafter \@firstoftwo
 \else \expandafter \@secondoftwo
 \fi
}%
\providecommand \@ifx [1]{%
 \ifx #1\expandafter \@firstoftwo
 \else \expandafter \@secondoftwo
 \fi
}%
\providecommand \natexlab [1]{#1}%
\providecommand \enquote  [1]{``#1''}%
\providecommand \bibnamefont  [1]{#1}%
\providecommand \bibfnamefont [1]{#1}%
\providecommand \citenamefont [1]{#1}%
\providecommand \href@noop [0]{\@secondoftwo}%
\providecommand \href [0]{\begingroup \@sanitize@url \@href}%
\providecommand \@href[1]{\@@startlink{#1}\@@href}%
\providecommand \@@href[1]{\endgroup#1\@@endlink}%
\providecommand \@sanitize@url [0]{\catcode `\\12\catcode `\$12\catcode
  `\&12\catcode `\#12\catcode `\^12\catcode `\_12\catcode `\%12\relax}%
\providecommand \@@startlink[1]{}%
\providecommand \@@endlink[0]{}%
\providecommand \url  [0]{\begingroup\@sanitize@url \@url }%
\providecommand \@url [1]{\endgroup\@href {#1}{\urlprefix }}%
\providecommand \urlprefix  [0]{URL }%
\providecommand \Eprint [0]{\href }%
\providecommand \doibase [0]{http://dx.doi.org/}%
\providecommand \selectlanguage [0]{\@gobble}%
\providecommand \bibinfo  [0]{\@secondoftwo}%
\providecommand \bibfield  [0]{\@secondoftwo}%
\providecommand \translation [1]{[#1]}%
\providecommand \BibitemOpen [0]{}%
\providecommand \bibitemStop [0]{}%
\providecommand \bibitemNoStop [0]{.\EOS\space}%
\providecommand \EOS [0]{\spacefactor3000\relax}%
\providecommand \BibitemShut  [1]{\csname bibitem#1\endcsname}%
\let\auto@bib@innerbib\@empty
\bibitem [{\citenamefont {{King}}\ \emph {et~al.}(1975)\citenamefont {{King}},
  \citenamefont {{Lasota}},\ and\ \citenamefont
  {{Kundt}}}]{1975PhRvD..12.3037K}%
  \BibitemOpen
  \bibfield  {author} {\bibinfo {author} {\bibfnamefont {A.~R.}\ \bibnamefont
  {{King}}}, \bibinfo {author} {\bibfnamefont {J.~P.}\ \bibnamefont
  {{Lasota}}}, \ and\ \bibinfo {author} {\bibfnamefont {W.}~\bibnamefont
  {{Kundt}}},\ }\href {\doibase 10.1103/PhysRevD.12.3037} {\bibfield  {journal}
  {\bibinfo  {journal} {\prd}\ }\textbf {\bibinfo {volume} {12}},\ \bibinfo
  {pages} {3037} (\bibinfo {year} {1975})}\BibitemShut {NoStop}%
\bibitem [{\citenamefont {{Bi{\v c}{\' a}k}}\ and\ \citenamefont
  {{Janis}}(1985)}]{1985MNRAS.212..899B}%
  \BibitemOpen
  \bibfield  {author} {\bibinfo {author} {\bibfnamefont {J.}~\bibnamefont
  {{Bi{\v c}{\' a}k}}}\ and\ \bibinfo {author} {\bibfnamefont {V.}~\bibnamefont
  {{Janis}}},\ }\href@noop {} {\bibfield  {journal} {\bibinfo  {journal}
  {\mnras}\ }\textbf {\bibinfo {volume} {212}},\ \bibinfo {pages} {899}
  (\bibinfo {year} {1985})}\BibitemShut {NoStop}%
\bibitem [{\citenamefont {{Bi{\v c}{\'a}k}}\ and\ \citenamefont
  {{Dvor{\'a}k}}(1980)}]{1980PhRvD..22.2933B}%
  \BibitemOpen
  \bibfield  {author} {\bibinfo {author} {\bibfnamefont {J.}~\bibnamefont
  {{Bi{\v c}{\'a}k}}}\ and\ \bibinfo {author} {\bibfnamefont {L.}~\bibnamefont
  {{Dvor{\'a}k}}},\ }\href {\doibase 10.1103/PhysRevD.22.2933} {\bibfield
  {journal} {\bibinfo  {journal} {\prd}\ }\textbf {\bibinfo {volume} {22}},\
  \bibinfo {pages} {2933} (\bibinfo {year} {1980})}\BibitemShut {NoStop}%
\bibitem [{\citenamefont {{Bi{\v c}{\'a}k}}\ and\ \citenamefont
  {{Karas}}(1989)}]{bicak1989}%
  \BibitemOpen
  \bibfield  {author} {\bibinfo {author} {\bibfnamefont {J.}~\bibnamefont
  {{Bi{\v c}{\'a}k}}}\ and\ \bibinfo {author} {\bibfnamefont {V.}~\bibnamefont
  {{Karas}}},\ }in\ \href@noop {} {\emph {\bibinfo {booktitle} {Proc. of the
  5th Marcel Grossman Meeting on General Relativity}}},\ \bibinfo {editor}
  {edited by\ \bibinfo {editor} {\bibfnamefont {D.~G.}\ \bibnamefont
  {{Blair}}}\ and\ \bibinfo {editor} {\bibfnamefont {M.~K.}\ \bibnamefont
  {{Buckingham}}}}\ (\bibinfo  {publisher} {World Scientific: Singapore},\
  \bibinfo {year} {1989})\ p.\ \bibinfo {pages} {1199}\BibitemShut {NoStop}%
\bibitem [{\citenamefont {{Karas}}\ and\ \citenamefont
  {{Vokrouhlick{\'y}}}(1991)}]{1991JMP....32..714K}%
  \BibitemOpen
  \bibfield  {author} {\bibinfo {author} {\bibfnamefont {V.}~\bibnamefont
  {{Karas}}}\ and\ \bibinfo {author} {\bibfnamefont {D.}~\bibnamefont
  {{Vokrouhlick{\'y}}}},\ }\href {\doibase 10.1063/1.529360} {\bibfield
  {journal} {\bibinfo  {journal} {Journal of Mathematical Physics}\ }\textbf
  {\bibinfo {volume} {32}},\ \bibinfo {pages} {714} (\bibinfo {year}
  {1991})}\BibitemShut {NoStop}%
\bibitem [{\citenamefont {{Karas}}\ and\ \citenamefont
  {{Budinov{\'a}}}(2000)}]{2000PhyS...61..253K}%
  \BibitemOpen
  \bibfield  {author} {\bibinfo {author} {\bibfnamefont {V.}~\bibnamefont
  {{Karas}}}\ and\ \bibinfo {author} {\bibfnamefont {Z.}~\bibnamefont
  {{Budinov{\'a}}}},\ }\href {\doibase 10.1238/Physica.Regular.061a00253}
  {\bibfield  {journal} {\bibinfo  {journal} {\physscr}\ }\textbf {\bibinfo
  {volume} {61}},\ \bibinfo {pages} {253} (\bibinfo {year} {2000})}\BibitemShut
  {NoStop}%
\bibitem [{\citenamefont {{Gibbons}}\ \emph {et~al.}(2014)\citenamefont
  {{Gibbons}}, \citenamefont {{Pang}},\ and\ \citenamefont
  {{Pope}}}]{2014PhRvD..89d4029G}%
  \BibitemOpen
  \bibfield  {author} {\bibinfo {author} {\bibfnamefont {G.~W.}\ \bibnamefont
  {{Gibbons}}}, \bibinfo {author} {\bibfnamefont {Y.}~\bibnamefont {{Pang}}}, \
  and\ \bibinfo {author} {\bibfnamefont {C.~N.}\ \bibnamefont {{Pope}}},\
  }\href {\doibase 10.1103/PhysRevD.89.044029} {\bibfield  {journal} {\bibinfo
  {journal} {\prd}\ }\textbf {\bibinfo {volume} {89}},\ \bibinfo {eid} {044029}
  (\bibinfo {year} {2014})}\BibitemShut {NoStop}%
\bibitem [{\citenamefont {{Chamblin}}\ \emph {et~al.}(1998)\citenamefont
  {{Chamblin}}, \citenamefont {{Emparan}},\ and\ \citenamefont
  {{Gibbons}}}]{1998PhRvD..58h4009C}%
  \BibitemOpen
  \bibfield  {author} {\bibinfo {author} {\bibfnamefont {A.}~\bibnamefont
  {{Chamblin}}}, \bibinfo {author} {\bibfnamefont {R.}~\bibnamefont
  {{Emparan}}}, \ and\ \bibinfo {author} {\bibfnamefont {G.~W.}\ \bibnamefont
  {{Gibbons}}},\ }\href {\doibase 10.1103/PhysRevD.58.084009} {\bibfield
  {journal} {\bibinfo  {journal} {\prd}\ }\textbf {\bibinfo {volume} {58}},\
  \bibinfo {eid} {084009} (\bibinfo {year} {1998})},\ \Eprint
  {http://arxiv.org/abs/hep-th/9806017} {hep-th/9806017} \BibitemShut {NoStop}%
\bibitem [{\citenamefont {{Ruffini}}(2004)}]{2004NCimB.119..785R}%
  \BibitemOpen
  \bibfield  {author} {\bibinfo {author} {\bibfnamefont {R.}~\bibnamefont
  {{Ruffini}}},\ }\href {\doibase 10.1393/ncb/i2004-10213-8} {\bibfield
  {journal} {\bibinfo  {journal} {Nuovo Cimento B Serie}\ }\textbf {\bibinfo
  {volume} {119}},\ \bibinfo {pages} {785} (\bibinfo {year} {2004})},\ \Eprint
  {http://arxiv.org/abs/astro-ph/0503439} {astro-ph/0503439} \BibitemShut
  {NoStop}%
\bibitem [{\citenamefont {{Bini}}\ \emph {et~al.}(2008)\citenamefont {{Bini}},
  \citenamefont {{Geralico}},\ and\ \citenamefont
  {{Ruffini}}}]{2008PhRvD..77f4020B}%
  \BibitemOpen
  \bibfield  {author} {\bibinfo {author} {\bibfnamefont {D.}~\bibnamefont
  {{Bini}}}, \bibinfo {author} {\bibfnamefont {A.}~\bibnamefont {{Geralico}}},
  \ and\ \bibinfo {author} {\bibfnamefont {R.}~\bibnamefont {{Ruffini}}},\
  }\href {\doibase 10.1103/PhysRevD.77.064020} {\bibfield  {journal} {\bibinfo
  {journal} {\prd}\ }\textbf {\bibinfo {volume} {77}},\ \bibinfo {eid} {064020}
  (\bibinfo {year} {2008})}\BibitemShut {NoStop}%
\bibitem [{\citenamefont {{Blandford}}\ and\ \citenamefont
  {{Znajek}}(1977)}]{bz77}%
  \BibitemOpen
  \bibfield  {author} {\bibinfo {author} {\bibfnamefont {R.~D.}\ \bibnamefont
  {{Blandford}}}\ and\ \bibinfo {author} {\bibfnamefont {R.~L.}\ \bibnamefont
  {{Znajek}}},\ }\href@noop {} {\bibfield  {journal} {\bibinfo  {journal}
  {\mnras}\ }\textbf {\bibinfo {volume} {179}},\ \bibinfo {pages} {433}
  (\bibinfo {year} {1977})}\BibitemShut {NoStop}%
\bibitem [{\citenamefont {{Penna}}(2014)}]{2014arXiv1403.0938P}%
  \BibitemOpen
  \bibfield  {author} {\bibinfo {author} {\bibfnamefont {R.~F.}\ \bibnamefont
  {{Penna}}},\ }\href@noop {} {\bibfield  {journal} {\bibinfo  {journal} {ArXiv
  e-prints}\ } (\bibinfo {year} {2014})},\ \Eprint
  {http://arxiv.org/abs/1403.0938} {arXiv:1403.0938 [astro-ph.HE]} \BibitemShut
  {NoStop}%
\bibitem [{\citenamefont {{Thorne}}\ \emph {et~al.}(1986)\citenamefont
  {{Thorne}}, \citenamefont {{Price}},\ and\ \citenamefont
  {{MacDonald}}}]{1986bhmp.book.....T}%
  \BibitemOpen
  \bibfield  {author} {\bibinfo {author} {\bibfnamefont {K.~S.}\ \bibnamefont
  {{Thorne}}}, \bibinfo {author} {\bibfnamefont {R.~H.}\ \bibnamefont
  {{Price}}}, \ and\ \bibinfo {author} {\bibfnamefont {D.~A.}\ \bibnamefont
  {{MacDonald}}},\ }\href@noop {} {\emph {\bibinfo {title} {Black Holes: The
  Membrane Paradigm}}}\ (\bibinfo  {publisher} {Yale University Press},\
  \bibinfo {year} {1986})\BibitemShut {NoStop}%
\bibitem [{\citenamefont {{Penna}}\ \emph {et~al.}(2013)\citenamefont
  {{Penna}}, \citenamefont {{Narayan}},\ and\ \citenamefont {{S{\c
  a}dowski}}}]{2013MNRAS.436.3741P}%
  \BibitemOpen
  \bibfield  {author} {\bibinfo {author} {\bibfnamefont {R.~F.}\ \bibnamefont
  {{Penna}}}, \bibinfo {author} {\bibfnamefont {R.}~\bibnamefont {{Narayan}}},
  \ and\ \bibinfo {author} {\bibfnamefont {A.}~\bibnamefont {{S{\c
  a}dowski}}},\ }\href {\doibase 10.1093/mnras/stt1860} {\bibfield  {journal}
  {\bibinfo  {journal} {\mnras}\ }\textbf {\bibinfo {volume} {436}},\ \bibinfo
  {pages} {3741} (\bibinfo {year} {2013})},\ \Eprint
  {http://arxiv.org/abs/1307.4752} {arXiv:1307.4752} \BibitemShut {NoStop}%
\bibitem [{\citenamefont {{Price}}(1991)}]{1991NYASA.631..235P}%
  \BibitemOpen
  \bibfield  {author} {\bibinfo {author} {\bibfnamefont {R.~H.}\ \bibnamefont
  {{Price}}},\ }\href {\doibase 10.1111/j.1749-6632.1991.tb52646.x} {\bibfield
  {journal} {\bibinfo  {journal} {Annals of the New York Academy of Sciences}\
  }\textbf {\bibinfo {volume} {631}},\ \bibinfo {pages} {235} (\bibinfo {year}
  {1991})}\BibitemShut {NoStop}%
\bibitem [{\citenamefont {{Bi{\v c}{\'a}k}}\ and\ \citenamefont
  {{Ledvinka}}(2000)}]{2000NCimB.115..739B}%
  \BibitemOpen
  \bibfield  {author} {\bibinfo {author} {\bibfnamefont {J.}~\bibnamefont
  {{Bi{\v c}{\'a}k}}}\ and\ \bibinfo {author} {\bibfnamefont {T.}~\bibnamefont
  {{Ledvinka}}},\ }\href@noop {} {\bibfield  {journal} {\bibinfo  {journal}
  {Nuovo Cimento B Serie}\ }\textbf {\bibinfo {volume} {115}},\ \bibinfo
  {pages} {739} (\bibinfo {year} {2000})},\ \Eprint
  {http://arxiv.org/abs/gr-qc/0012006} {gr-qc/0012006} \BibitemShut {NoStop}%
\bibitem [{\citenamefont {{Bi{\v c}{\'a}k}}\ \emph {et~al.}(2007)\citenamefont
  {{Bi{\v c}{\'a}k}}, \citenamefont {{Karas}},\ and\ \citenamefont
  {{Ledvinka}}}]{2007IAUS..238..139B}%
  \BibitemOpen
  \bibfield  {author} {\bibinfo {author} {\bibfnamefont {J.}~\bibnamefont
  {{Bi{\v c}{\'a}k}}}, \bibinfo {author} {\bibfnamefont {V.}~\bibnamefont
  {{Karas}}}, \ and\ \bibinfo {author} {\bibfnamefont {T.}~\bibnamefont
  {{Ledvinka}}},\ }in\ \href {\doibase 10.1017/S1743921307004851} {\emph
  {\bibinfo {booktitle} {IAU Symposium}}},\ \bibinfo {series} {IAU Symposium},
  Vol.\ \bibinfo {volume} {238},\ \bibinfo {editor} {edited by\ \bibinfo
  {editor} {\bibfnamefont {V.}~\bibnamefont {{Karas}}}\ and\ \bibinfo {editor}
  {\bibfnamefont {G.}~\bibnamefont {{Matt}}}}\ (\bibinfo {year} {2007})\ pp.\
  \bibinfo {pages} {139--144},\ \Eprint {http://arxiv.org/abs/astro-ph/0610841}
  {astro-ph/0610841} \BibitemShut {NoStop}%
\bibitem [{\citenamefont {{Komissarov}}\ and\ \citenamefont
  {{McKinney}}(2007)}]{2007MNRAS.377L..49K}%
  \BibitemOpen
  \bibfield  {author} {\bibinfo {author} {\bibfnamefont {S.~S.}\ \bibnamefont
  {{Komissarov}}}\ and\ \bibinfo {author} {\bibfnamefont {J.~C.}\ \bibnamefont
  {{McKinney}}},\ }\href {\doibase 10.1111/j.1745-3933.2007.00301.x} {\bibfield
   {journal} {\bibinfo  {journal} {\mnras}\ }\textbf {\bibinfo {volume}
  {377}},\ \bibinfo {pages} {L49} (\bibinfo {year} {2007})},\ \Eprint
  {http://arxiv.org/abs/astro-ph/0702269} {astro-ph/0702269} \BibitemShut
  {NoStop}%
\bibitem [{\citenamefont {{Bardeen}}\ \emph {et~al.}(1972)\citenamefont
  {{Bardeen}}, \citenamefont {{Press}},\ and\ \citenamefont
  {{Teukolsky}}}]{1972ApJ...178..347B}%
  \BibitemOpen
  \bibfield  {author} {\bibinfo {author} {\bibfnamefont {J.~M.}\ \bibnamefont
  {{Bardeen}}}, \bibinfo {author} {\bibfnamefont {W.~H.}\ \bibnamefont
  {{Press}}}, \ and\ \bibinfo {author} {\bibfnamefont {S.~A.}\ \bibnamefont
  {{Teukolsky}}},\ }\href {\doibase 10.1086/151796} {\bibfield  {journal}
  {\bibinfo  {journal} {\apj}\ }\textbf {\bibinfo {volume} {178}},\ \bibinfo
  {pages} {347} (\bibinfo {year} {1972})}\BibitemShut {NoStop}%
\bibitem [{\citenamefont {{Maldacena}}(2003)}]{2003JHEP...04..021M}%
  \BibitemOpen
  \bibfield  {author} {\bibinfo {author} {\bibfnamefont {J.}~\bibnamefont
  {{Maldacena}}},\ }\href {\doibase 10.1088/1126-6708/2003/04/021} {\bibfield
  {journal} {\bibinfo  {journal} {Journal of High Energy Physics}\ }\textbf
  {\bibinfo {volume} {4}},\ \bibinfo {eid} {021} (\bibinfo {year} {2003})},\
  \Eprint {http://arxiv.org/abs/hep-th/0106112} {hep-th/0106112} \BibitemShut
  {NoStop}%
\bibitem [{\citenamefont {{Balasubramanian}}\ and\ \citenamefont
  {{Ross}}(2000)}]{2000PhRvD..61d4007B}%
  \BibitemOpen
  \bibfield  {author} {\bibinfo {author} {\bibfnamefont {V.}~\bibnamefont
  {{Balasubramanian}}}\ and\ \bibinfo {author} {\bibfnamefont {S.~F.}\
  \bibnamefont {{Ross}}},\ }\href {\doibase 10.1103/PhysRevD.61.044007}
  {\bibfield  {journal} {\bibinfo  {journal} {\prd}\ }\textbf {\bibinfo
  {volume} {61}},\ \bibinfo {eid} {044007} (\bibinfo {year} {2000})},\ \Eprint
  {http://arxiv.org/abs/hep-th/9906226} {hep-th/9906226} \BibitemShut {NoStop}%
\bibitem [{\citenamefont {{Louko}}\ \emph {et~al.}(2000)\citenamefont
  {{Louko}}, \citenamefont {{Marolf}},\ and\ \citenamefont
  {{Ross}}}]{2000PhRvD..62d4041L}%
  \BibitemOpen
  \bibfield  {author} {\bibinfo {author} {\bibfnamefont {J.}~\bibnamefont
  {{Louko}}}, \bibinfo {author} {\bibfnamefont {D.}~\bibnamefont {{Marolf}}}, \
  and\ \bibinfo {author} {\bibfnamefont {S.~F.}\ \bibnamefont {{Ross}}},\
  }\href {\doibase 10.1103/PhysRevD.62.044041} {\bibfield  {journal} {\bibinfo
  {journal} {\prd}\ }\textbf {\bibinfo {volume} {62}},\ \bibinfo {eid} {044041}
  (\bibinfo {year} {2000})},\ \Eprint {http://arxiv.org/abs/hep-th/0002111}
  {hep-th/0002111} \BibitemShut {NoStop}%
\bibitem [{\citenamefont {{Kraus}}\ \emph {et~al.}(2003)\citenamefont
  {{Kraus}}, \citenamefont {{Ooguri}},\ and\ \citenamefont
  {{Shenker}}}]{2003PhRvD..67l4022K}%
  \BibitemOpen
  \bibfield  {author} {\bibinfo {author} {\bibfnamefont {P.}~\bibnamefont
  {{Kraus}}}, \bibinfo {author} {\bibfnamefont {H.}~\bibnamefont {{Ooguri}}}, \
  and\ \bibinfo {author} {\bibfnamefont {S.}~\bibnamefont {{Shenker}}},\ }\href
  {\doibase 10.1103/PhysRevD.67.124022} {\bibfield  {journal} {\bibinfo
  {journal} {\prd}\ }\textbf {\bibinfo {volume} {67}},\ \bibinfo {eid} {124022}
  (\bibinfo {year} {2003})},\ \Eprint {http://arxiv.org/abs/hep-th/0212277}
  {hep-th/0212277} \BibitemShut {NoStop}%
\bibitem [{\citenamefont {{Hubeny}}\ and\ \citenamefont
  {{Maxfield}}(2014)}]{2014JHEP...03..097H}%
  \BibitemOpen
  \bibfield  {author} {\bibinfo {author} {\bibfnamefont {V.~E.}\ \bibnamefont
  {{Hubeny}}}\ and\ \bibinfo {author} {\bibfnamefont {H.}~\bibnamefont
  {{Maxfield}}},\ }\href {\doibase 10.1007/JHEP03(2014)097} {\bibfield
  {journal} {\bibinfo  {journal} {Journal of High Energy Physics}\ }\textbf
  {\bibinfo {volume} {3}},\ \bibinfo {pages} {97} (\bibinfo {year} {2014})},\
  \Eprint {http://arxiv.org/abs/1312.6887} {arXiv:1312.6887 [hep-th]}
  \BibitemShut {NoStop}%
\bibitem [{\citenamefont {{van Raamsdonk}}(2010)}]{2010GReGr..42.2323V}%
  \BibitemOpen
  \bibfield  {author} {\bibinfo {author} {\bibfnamefont {M.}~\bibnamefont {{van
  Raamsdonk}}},\ }\href {\doibase 10.1007/s10714-010-1034-0} {\bibfield
  {journal} {\bibinfo  {journal} {General Relativity and Gravitation}\ }\textbf
  {\bibinfo {volume} {42}},\ \bibinfo {pages} {2323} (\bibinfo {year}
  {2010})},\ \Eprint {http://arxiv.org/abs/1005.3035} {arXiv:1005.3035
  [hep-th]} \BibitemShut {NoStop}%
\bibitem [{\citenamefont {Andrade}\ \emph {et~al.}(2014)\citenamefont
  {Andrade}, \citenamefont {Fischetti}, \citenamefont {Marolf}, \citenamefont
  {Ross},\ and\ \citenamefont {Rozali}}]{Andrade:2013rra}%
  \BibitemOpen
  \bibfield  {author} {\bibinfo {author} {\bibfnamefont {T.}~\bibnamefont
  {Andrade}}, \bibinfo {author} {\bibfnamefont {S.}~\bibnamefont {Fischetti}},
  \bibinfo {author} {\bibfnamefont {D.}~\bibnamefont {Marolf}}, \bibinfo
  {author} {\bibfnamefont {S.~F.}\ \bibnamefont {Ross}}, \ and\ \bibinfo
  {author} {\bibfnamefont {M.}~\bibnamefont {Rozali}},\ }\href {\doibase
  10.1007/JHEP04(2014)023} {\bibfield  {journal} {\bibinfo  {journal} {JHEP}\
  }\textbf {\bibinfo {volume} {1404}},\ \bibinfo {pages} {023} (\bibinfo {year}
  {2014})},\ \Eprint {http://arxiv.org/abs/1312.2839} {arXiv:1312.2839
  [hep-th]} \BibitemShut {NoStop}%
\bibitem [{\citenamefont {Leichenauer}(2014)}]{Leichenauer:2014nxa}%
  \BibitemOpen
  \bibfield  {author} {\bibinfo {author} {\bibfnamefont {S.}~\bibnamefont
  {Leichenauer}},\ }\href@noop {} {\bibfield  {journal} {\bibinfo  {journal}
  {{arXiv}}\ } (\bibinfo {year} {2014})},\ \Eprint
  {http://arxiv.org/abs/1405.7365} {arXiv:1405.7365 [hep-th]} \BibitemShut
  {NoStop}%
\bibitem [{\citenamefont {{Hartle}}\ and\ \citenamefont
  {{Hawking}}(1976)}]{1976PhRvD..13.2188H}%
  \BibitemOpen
  \bibfield  {author} {\bibinfo {author} {\bibfnamefont {J.~B.}\ \bibnamefont
  {{Hartle}}}\ and\ \bibinfo {author} {\bibfnamefont {S.~W.}\ \bibnamefont
  {{Hawking}}},\ }\href {\doibase 10.1103/PhysRevD.13.2188} {\bibfield
  {journal} {\bibinfo  {journal} {\prd}\ }\textbf {\bibinfo {volume} {13}},\
  \bibinfo {pages} {2188} (\bibinfo {year} {1976})}\BibitemShut {NoStop}%
\bibitem [{\citenamefont {{Israel}}(1976)}]{1976PhLA...57..107I}%
  \BibitemOpen
  \bibfield  {author} {\bibinfo {author} {\bibfnamefont {W.}~\bibnamefont
  {{Israel}}},\ }\href {\doibase 10.1016/0375-9601(76)90178-X} {\bibfield
  {journal} {\bibinfo  {journal} {Physics Letters A}\ }\textbf {\bibinfo
  {volume} {57}},\ \bibinfo {pages} {107} (\bibinfo {year} {1976})}\BibitemShut
  {NoStop}%
\bibitem [{\citenamefont {Kubo}(1957)}]{doi:10.1143/JPSJ.12.570}%
  \BibitemOpen
  \bibfield  {author} {\bibinfo {author} {\bibfnamefont {R.}~\bibnamefont
  {Kubo}},\ }\href {\doibase 10.1143/JPSJ.12.570} {\bibfield  {journal}
  {\bibinfo  {journal} {Journal of the Physical Society of Japan}\ }\textbf
  {\bibinfo {volume} {12}},\ \bibinfo {pages} {570} (\bibinfo {year} {1957})},\
  \Eprint {http://arxiv.org/abs/http://dx.doi.org/10.1143/JPSJ.12.570}
  {http://dx.doi.org/10.1143/JPSJ.12.570} \BibitemShut {NoStop}%
\bibitem [{\citenamefont {Martin}\ and\ \citenamefont
  {Schwinger}(1959)}]{PhysRev.115.1342}%
  \BibitemOpen
  \bibfield  {author} {\bibinfo {author} {\bibfnamefont {P.~C.}\ \bibnamefont
  {Martin}}\ and\ \bibinfo {author} {\bibfnamefont {J.}~\bibnamefont
  {Schwinger}},\ }\href {\doibase 10.1103/PhysRev.115.1342} {\bibfield
  {journal} {\bibinfo  {journal} {Phys. Rev.}\ }\textbf {\bibinfo {volume}
  {115}},\ \bibinfo {pages} {1342} (\bibinfo {year} {1959})}\BibitemShut
  {NoStop}%
\bibitem [{\citenamefont {{Takahashi}}\ and\ \citenamefont
  {{Umezawa}}(1996)}]{1996IJMPB..10.1755T}%
  \BibitemOpen
  \bibfield  {author} {\bibinfo {author} {\bibfnamefont {Y.}~\bibnamefont
  {{Takahashi}}}\ and\ \bibinfo {author} {\bibfnamefont {H.}~\bibnamefont
  {{Umezawa}}},\ }\href {\doibase 10.1142/S0217979296000817} {\bibfield
  {journal} {\bibinfo  {journal} {International Journal of Modern Physics B}\
  }\textbf {\bibinfo {volume} {10}},\ \bibinfo {pages} {1755} (\bibinfo {year}
  {1996})}\BibitemShut {NoStop}%
\bibitem [{\citenamefont {{Balachandran}}\ \emph {et~al.}(1995)\citenamefont
  {{Balachandran}}, \citenamefont {{Chandar}},\ and\ \citenamefont
  {{Ercolessi}}}]{1995IJMPA..10.1969B}%
  \BibitemOpen
  \bibfield  {author} {\bibinfo {author} {\bibfnamefont {A.~P.}\ \bibnamefont
  {{Balachandran}}}, \bibinfo {author} {\bibfnamefont {L.}~\bibnamefont
  {{Chandar}}}, \ and\ \bibinfo {author} {\bibfnamefont {E.}~\bibnamefont
  {{Ercolessi}}},\ }\href {\doibase 10.1142/S0217751X95000966} {\bibfield
  {journal} {\bibinfo  {journal} {International Journal of Modern Physics A}\
  }\textbf {\bibinfo {volume} {10}},\ \bibinfo {pages} {1969} (\bibinfo {year}
  {1995})},\ \Eprint {http://arxiv.org/abs/hep-th/9411164} {hep-th/9411164}
  \BibitemShut {NoStop}%
\bibitem [{\citenamefont {{Balachandran}}\ \emph {et~al.}(1996)\citenamefont
  {{Balachandran}}, \citenamefont {{Momen}},\ and\ \citenamefont
  {{Chandar}}}]{1996NuPhB.461..581B}%
  \BibitemOpen
  \bibfield  {author} {\bibinfo {author} {\bibfnamefont {A.~P.}\ \bibnamefont
  {{Balachandran}}}, \bibinfo {author} {\bibfnamefont {A.}~\bibnamefont
  {{Momen}}}, \ and\ \bibinfo {author} {\bibfnamefont {L.}~\bibnamefont
  {{Chandar}}},\ }\href {\doibase 10.1016/0550-3213(95)00622-2} {\bibfield
  {journal} {\bibinfo  {journal} {Nuclear Physics B}\ }\textbf {\bibinfo
  {volume} {461}},\ \bibinfo {pages} {581} (\bibinfo {year} {1996})},\ \Eprint
  {http://arxiv.org/abs/gr-qc/9412019} {gr-qc/9412019} \BibitemShut {NoStop}%
\bibitem [{\citenamefont {{Buividovich}}\ and\ \citenamefont
  {{Polikarpov}}(2008)}]{2008PhLB..670..141B}%
  \BibitemOpen
  \bibfield  {author} {\bibinfo {author} {\bibfnamefont {P.~V.}\ \bibnamefont
  {{Buividovich}}}\ and\ \bibinfo {author} {\bibfnamefont {M.~I.}\ \bibnamefont
  {{Polikarpov}}},\ }\href {\doibase 10.1016/j.physletb.2008.10.032} {\bibfield
   {journal} {\bibinfo  {journal} {Physics Letters B}\ }\textbf {\bibinfo
  {volume} {670}},\ \bibinfo {pages} {141} (\bibinfo {year} {2008})},\ \Eprint
  {http://arxiv.org/abs/0806.3376} {arXiv:0806.3376 [hep-th]} \BibitemShut
  {NoStop}%
\bibitem [{\citenamefont {{Donnelly}}(2012)}]{2012PhRvD..85h5004D}%
  \BibitemOpen
  \bibfield  {author} {\bibinfo {author} {\bibfnamefont {W.}~\bibnamefont
  {{Donnelly}}},\ }\href {\doibase 10.1103/PhysRevD.85.085004} {\bibfield
  {journal} {\bibinfo  {journal} {\prd}\ }\textbf {\bibinfo {volume} {85}},\
  \bibinfo {eid} {085004} (\bibinfo {year} {2012})},\ \Eprint
  {http://arxiv.org/abs/1109.0036} {arXiv:1109.0036 [hep-th]} \BibitemShut
  {NoStop}%
\bibitem [{\citenamefont {{Kay}}\ and\ \citenamefont
  {{Wald}}(1991)}]{1991PhR...207...49K}%
  \BibitemOpen
  \bibfield  {author} {\bibinfo {author} {\bibfnamefont {B.~S.}\ \bibnamefont
  {{Kay}}}\ and\ \bibinfo {author} {\bibfnamefont {R.~M.}\ \bibnamefont
  {{Wald}}},\ }\href {\doibase 10.1016/0370-1573(91)90015-E} {\bibfield
  {journal} {\bibinfo  {journal} {\physrep}\ }\textbf {\bibinfo {volume}
  {207}},\ \bibinfo {pages} {49} (\bibinfo {year} {1991})}\BibitemShut
  {NoStop}%
\bibitem [{Note1()}]{Note1}%
  \BibitemOpen
  \bibinfo {note} {This is related to the fact that the metric does not have a
  globally defined timelike Killing vector}\BibitemShut {NoStop}%
\bibitem [{\citenamefont {Duffy}\ and\ \citenamefont
  {Ottewill}(2008)}]{Duffy:2005mz}%
  \BibitemOpen
  \bibfield  {author} {\bibinfo {author} {\bibfnamefont {G.}~\bibnamefont
  {Duffy}}\ and\ \bibinfo {author} {\bibfnamefont {A.~C.}\ \bibnamefont
  {Ottewill}},\ }\href {\doibase 10.1103/PhysRevD.77.024007} {\bibfield
  {journal} {\bibinfo  {journal} {Phys.Rev.}\ }\textbf {\bibinfo {volume}
  {D77}},\ \bibinfo {pages} {024007} (\bibinfo {year} {2008})},\ \Eprint
  {http://arxiv.org/abs/gr-qc/0507116} {arXiv:gr-qc/0507116 [gr-qc]}
  \BibitemShut {NoStop}%
\bibitem [{\citenamefont {Frolov}\ and\ \citenamefont
  {Thorne}(1989)}]{Frolov:1989jh}%
  \BibitemOpen
  \bibfield  {author} {\bibinfo {author} {\bibfnamefont {V.~P.}\ \bibnamefont
  {Frolov}}\ and\ \bibinfo {author} {\bibfnamefont {K.}~\bibnamefont
  {Thorne}},\ }\href {\doibase 10.1103/PhysRevD.39.2125} {\bibfield  {journal}
  {\bibinfo  {journal} {Phys.Rev.}\ }\textbf {\bibinfo {volume} {D39}},\
  \bibinfo {pages} {2125} (\bibinfo {year} {1989})}\BibitemShut {NoStop}%
\bibitem [{\citenamefont {Ojima}(1981)}]{Ojima19811}%
  \BibitemOpen
  \bibfield  {author} {\bibinfo {author} {\bibfnamefont {I.}~\bibnamefont
  {Ojima}},\ }\href {\doibase http://dx.doi.org/10.1016/0003-4916(81)90058-0}
  {\bibfield  {journal} {\bibinfo  {journal} {Annals of Physics}\ }\textbf
  {\bibinfo {volume} {137}},\ \bibinfo {pages} {1 } (\bibinfo {year}
  {1981})}\BibitemShut {NoStop}%
\bibitem [{\citenamefont {Deser}\ and\ \citenamefont
  {Levin}(1997)}]{Deser:1997ri}%
  \BibitemOpen
  \bibfield  {author} {\bibinfo {author} {\bibfnamefont {S.}~\bibnamefont
  {Deser}}\ and\ \bibinfo {author} {\bibfnamefont {O.}~\bibnamefont {Levin}},\
  }\href {\doibase 10.1088/0264-9381/14/9/003} {\bibfield  {journal} {\bibinfo
  {journal} {Class.Quant.Grav.}\ }\textbf {\bibinfo {volume} {14}},\ \bibinfo
  {pages} {L163} (\bibinfo {year} {1997})},\ \Eprint
  {http://arxiv.org/abs/gr-qc/9706018} {arXiv:gr-qc/9706018 [gr-qc]}
  \BibitemShut {NoStop}%
\bibitem [{\citenamefont {Parikh}\ and\ \citenamefont
  {Samantray}(2012)}]{Parikh:2012kg}%
  \BibitemOpen
  \bibfield  {author} {\bibinfo {author} {\bibfnamefont {M.}~\bibnamefont
  {Parikh}}\ and\ \bibinfo {author} {\bibfnamefont {P.}~\bibnamefont
  {Samantray}},\ }\href@noop {} {\bibfield  {journal} {\bibinfo  {journal}
  {{arXiv}}\ } (\bibinfo {year} {2012})},\ \Eprint
  {http://arxiv.org/abs/1211.7370} {arXiv:1211.7370 [hep-th]} \BibitemShut
  {NoStop}%
\bibitem [{\citenamefont {{McClintock}}\ \emph {et~al.}(2006)\citenamefont
  {{McClintock}}, \citenamefont {{Shafee}}, \citenamefont {{Narayan}},
  \citenamefont {{Remillard}}, \citenamefont {{Davis}},\ and\ \citenamefont
  {{Li}}}]{2006ApJ...652..518M}%
  \BibitemOpen
  \bibfield  {author} {\bibinfo {author} {\bibfnamefont {J.~E.}\ \bibnamefont
  {{McClintock}}}, \bibinfo {author} {\bibfnamefont {R.}~\bibnamefont
  {{Shafee}}}, \bibinfo {author} {\bibfnamefont {R.}~\bibnamefont {{Narayan}}},
  \bibinfo {author} {\bibfnamefont {R.~A.}\ \bibnamefont {{Remillard}}},
  \bibinfo {author} {\bibfnamefont {S.~W.}\ \bibnamefont {{Davis}}}, \ and\
  \bibinfo {author} {\bibfnamefont {L.-X.}\ \bibnamefont {{Li}}},\ }\href
  {\doibase 10.1086/508457} {\bibfield  {journal} {\bibinfo  {journal} {\apj}\
  }\textbf {\bibinfo {volume} {652}},\ \bibinfo {pages} {518} (\bibinfo {year}
  {2006})},\ \Eprint {http://arxiv.org/abs/arXiv:astro-ph/0606076}
  {arXiv:astro-ph/0606076} \BibitemShut {NoStop}%
\bibitem [{\citenamefont {{Gou}}\ \emph {et~al.}(2011)\citenamefont {{Gou}},
  \citenamefont {{McClintock}}, \citenamefont {{Reid}}, \citenamefont
  {{Orosz}}, \citenamefont {{Steiner}}, \citenamefont {{Narayan}},
  \citenamefont {{Xiang}}, \citenamefont {{Remillard}}, \citenamefont
  {{Arnaud}},\ and\ \citenamefont {{Davis}}}]{2011ApJ...742...85G}%
  \BibitemOpen
  \bibfield  {author} {\bibinfo {author} {\bibfnamefont {L.}~\bibnamefont
  {{Gou}}}, \bibinfo {author} {\bibfnamefont {J.~E.}\ \bibnamefont
  {{McClintock}}}, \bibinfo {author} {\bibfnamefont {M.~J.}\ \bibnamefont
  {{Reid}}}, \bibinfo {author} {\bibfnamefont {J.~A.}\ \bibnamefont {{Orosz}}},
  \bibinfo {author} {\bibfnamefont {J.~F.}\ \bibnamefont {{Steiner}}}, \bibinfo
  {author} {\bibfnamefont {R.}~\bibnamefont {{Narayan}}}, \bibinfo {author}
  {\bibfnamefont {J.}~\bibnamefont {{Xiang}}}, \bibinfo {author} {\bibfnamefont
  {R.~A.}\ \bibnamefont {{Remillard}}}, \bibinfo {author} {\bibfnamefont
  {K.~A.}\ \bibnamefont {{Arnaud}}}, \ and\ \bibinfo {author} {\bibfnamefont
  {S.~W.}\ \bibnamefont {{Davis}}},\ }\href {\doibase
  10.1088/0004-637X/742/2/85} {\bibfield  {journal} {\bibinfo  {journal}
  {\apj}\ }\textbf {\bibinfo {volume} {742}},\ \bibinfo {eid} {85} (\bibinfo
  {year} {2011})},\ \Eprint {http://arxiv.org/abs/1106.3690} {arXiv:1106.3690}
  \BibitemShut {NoStop}%
\bibitem [{\citenamefont {{McClintock}}\ \emph {et~al.}(2013)\citenamefont
  {{McClintock}}, \citenamefont {{Narayan}},\ and\ \citenamefont
  {{Steiner}}}]{2013SSRv..tmp...73M}%
  \BibitemOpen
  \bibfield  {author} {\bibinfo {author} {\bibfnamefont {J.~E.}\ \bibnamefont
  {{McClintock}}}, \bibinfo {author} {\bibfnamefont {R.}~\bibnamefont
  {{Narayan}}}, \ and\ \bibinfo {author} {\bibfnamefont {J.~F.}\ \bibnamefont
  {{Steiner}}},\ }\href {\doibase 10.1007/s11214-013-0003-9} {\bibfield
  {journal} {\bibinfo  {journal} {\ssr}\ } (\bibinfo {year} {2013}),\
  10.1007/s11214-013-0003-9},\ \Eprint {http://arxiv.org/abs/1303.1583}
  {arXiv:1303.1583} \BibitemShut {NoStop}%
\bibitem [{\citenamefont {{Parikh}}\ and\ \citenamefont
  {{Wilczek}}(1998)}]{1998PhRvD..58f4011P}%
  \BibitemOpen
  \bibfield  {author} {\bibinfo {author} {\bibfnamefont {M.~K.}\ \bibnamefont
  {{Parikh}}}\ and\ \bibinfo {author} {\bibfnamefont {F.}~\bibnamefont
  {{Wilczek}}},\ }\href {\doibase 10.1103/PhysRevD.58.064011} {\bibfield
  {journal} {\bibinfo  {journal} {\prd}\ }\textbf {\bibinfo {volume} {58}},\
  \bibinfo {eid} {064011} (\bibinfo {year} {1998})},\ \Eprint
  {http://arxiv.org/abs/gr-qc/9712077} {gr-qc/9712077} \BibitemShut {NoStop}%
\end{thebibliography}%

\end{document}